# REVIEW

# MATERIALS SCIENCE

## Polar or nonpolar? That is not the question for perovskite solar cells

Boyuan Huang,[a,b] Zhenghao Liu,[c] Changwei Wu,[c] Yuan Zhang,[c] Jinjin Zhao,[d] Xiao Wang,[c,*] Jiangyu Li [a,b,*]

a   Department of Materials Science and Engineering, Southern University of Science and Technology, Shenzhen 518055, China;

b   Guangdong Key laboratory of Functional Oxide Materials and Devices, Southern University of Science and Technology, Shenzhen 518055, China;

c   Shenzhen Key Laboratory of Nanobiomechanics, Shenzhen Institutes of Advanced Technology, Chinese Academy of Science, Shenzhen 518055, China;

d   School of Materials Science and Engineering, Key Laboratory of Smart Materials and Structures Mechanics of Hebei Province, Shijiazhuang Tiedao University, Shijiazhuang 050043, China

***Corresponding authors.** E-mails: xiao.wang@siat.ac.cn; lijy@sustech.edu.cn

## ABSTRACT

Perovskite solar cell (PSC) is one of the most promising next generation photovoltaic technologies, and there are considerable interests in the role of possible polarization of organic-inorganic halide perovskites (OIHPs) in photovoltaic conversion. The polarity of OIHPs, however, is still hotly debated. In this review, we examine recent literature on the polarity of OIHPs from both theoretical and experimental points of view, and argue that they can be both polar and nonpolar, depending on compositions, processing, and environments. Implications of OIHP polarity to photovoltaic conversion is also discussed, and effort in answering these questions continues to render us new insights. In the future, integrating local scanning probe with global macroscopic measurements in-situ will provide invaluable microscopic insight into the intriguing macroscopic phenomena, while synchrotron diffractions and scanning transmission electron microscopy on more stable samples may ultimately settle the debate.

**INTRODUCTION**

Ever since the spectacular rise of perovskite solar cells (PSCs), there have been suggestions on possible roles of ferroelectric polarization in their photovoltaic conversion. Perovskite materials, particularly oxides, are often ferroelectric, and early theoretical calculations indicated that polarization in organic-inorganic halide perovskites (OIHPs) may help charge separation and facilitate carrier transport [1]. Nevertheless, the ferroelectricity of OIHPs has not been firmly established experimentally. In fact, the possible polarity of OIHPs is still hotly debated [2,3], and there are considerable theoretical and experimental evidences supporting either points of views [4], which is summarized in **Table 1**. As shown in **Fig. 1**, both nonpolar I4/mcm (**Fig. 1a**) and polar I4cm (**Fig. 1b**) space groups are possible for $CH_3NH_3PbI_3$ (MAPI) [5-7], and the structural difference is very subtle, making it difficult to differentiate by conventional structural characterization techniques such as diffractions. Indeed, the structure details of MAPI have not been fully resolved, and the poor stability of the materials makes the problem even worse. In this review, we examine recent literature on the polarity of OIHPs, and argue that they can be both polar and nonpolar, depending on compositions, processing, and environments. Implications to photovoltaic conversion are also discussed, especially hysteresis.

**Table 1.** Literature survey on the polarity of OIHPs.

| Technique | Non-polar I4/mcm | Polar I4cm | Noncommittal |
| --- | --- | --- | --- |
| X-ray and neutron diffractions | Refs [5,42-45] | Refs [6,41] | Ref. [40] |
| Optic SHG | Refs [12,19,62,67] | Refs [76,93] | |
| Macroscopic measurements | Refs [59,60,64] | Refs [48,50,76,90,101] | Ref. [75] |
| Microscopic PFM | Refs [8,63,65,68,91] | Refs [46,47,49-58,83] | Refs [61,80,81,104] |
| TEM | | Ref. [73] | Ref. [70] |
| DFT and MD simulations | Refs [12,19,66] | Ref. [92] | Refs [8,21] |

**THEORETICAL CONSIDERATIONS**

Different from traditional perovskite, the component at A site in OIHP is positioned by a molecule-type ion, which may have intrinsic dipole and induce the deformation of the octahedron framework caused by the interatomic hydrogen bond. Therefore, the apparent polarization of OIHP is the collective polarization of each unit impacted by the orientation of the A-site molecule. In the case of MAPI, the major structural difference between the polar I4cm and nonpolar I4/mcm phases is the orientation of MA cations, which have an intrinsic dipole of ~2.3 D [1]. In the polar phase, the C-N dipole shows a 'head-tail' alignment along the c axis and displays a large polarization of several $\mu C / cm^2$ [8-11]. While in the nonpolar counterpart, due to the space group symmetry, each MA cation is usually described with partial occupancies with 4 identical positions and thus exhibits no net polarization [12]. Nevertheless, the orientation of MA cation can distort the neighboring iodides from their centrosymmetric positions, leading to the ferroelectricity [13].

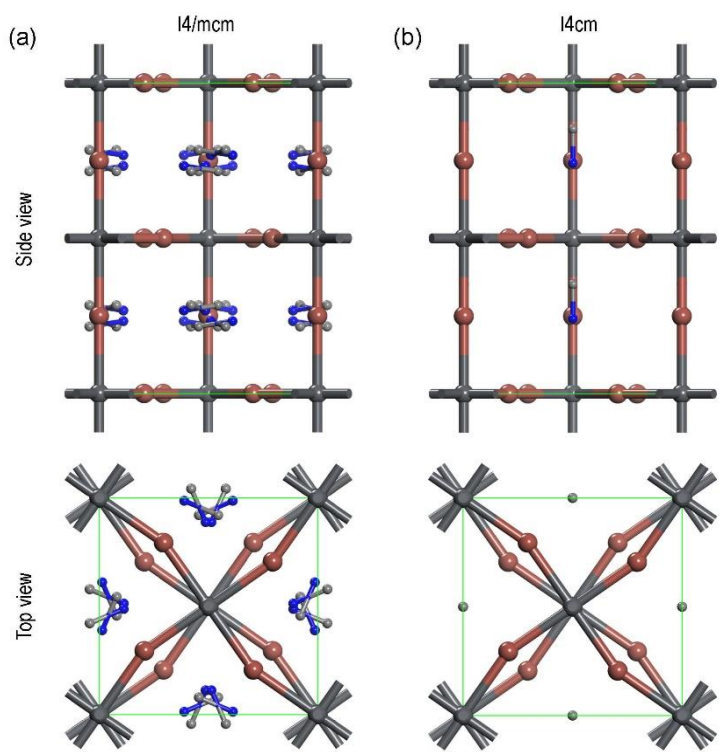

**Figure 1.** Schematic lattice of nonpolar I4/mcm (a) and polar I4cm (b) space groups for MAPI from side and top views. The hydrogen atoms are hidden for simplicity.

So far, the optimal orientation of MA in MAPbI$_3$ bulk has still been debated due to theoretical models and methods. Walsh et al. calculated the total energy when MA arrays along <100>, <110> and <111> directions in the cubic phase and found that <100> is the most stable orientation with energy difference less than 15 meV per atom [14]. Ven et al. calculated the full energy landscape for rigid-body rotations and translations of MA in the cubic phase and also reached the same conclusion. They revealed that the preferential orientation is attributed to the strong N-H···I interactions between MA and the Pb-I framework along <100> direction [15]. However, others came to the different conclusions. Vashishta et al. used a cubic symmetry-assisted analysis and found that the prominent orientation of MA is the crystalline <110> directions, rather than the <100> and <111> directions [16]. Zhang et al. studied the MA orientation using the swarm intelligence-based structure prediction method combined with DFT calculations, but they found that <012> orientation is the most stable one instead of the aforementioned directions [17].

Despite the puzzled optimal orientation, it reaches an agreement that the orientation of MA just tunes the strength and direction of the hydrogen bond between MA$^+$ and I$^-$ that is rather weak (~ 0.09 eV / cation) [18], there is only slight energy difference (< 0.1 eV / unit) between the two phases and phase transition barrier is quite small as well (about 0.2 eV / unit) [9]. Such tiny difference makes it easily accessible for the transition between the polar and nonpolar phases at room temperature [19,20]. Furthermore, the subtlety between the two phases also makes the debate regarding polar and nonpolar nature of OIHPs noticeably depending on method, model, size, and time-scale in ab initio calculations [9,21]. Indeed, it should be stressed that we should not only focus on the origin of the polarity in its primitive cell, but also the long-range dynamics of the MA cations in a wider vision. Ab initio molecular simulation is a versatile method that can consider more operational conditions (such as temperature, long-range dynamics etc.) with accuracy. The random order of MA and the phase transition between the two phases have been tracked, usually indicating an antiferroelectric nature of tetragonal OIHPs [19,20,22].

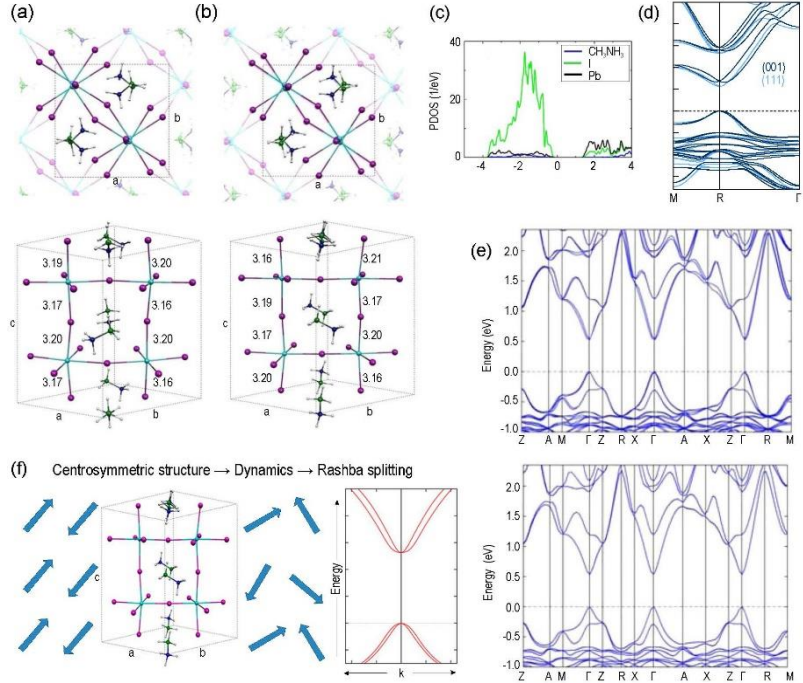

**Figure 2.** Theoretical studies on photovoltaic properties of polar and nonpolar MAPI in bulk phase. The top and side views of (a) polar and (b) nonpolar phases of MAPI relying on the orientation of MA cation. The Pb, I, C, N, H atoms are colored in light blue, pink, blue, green and white. (c) The projected density of states of MAPI. The Pb and I instead of MA group mainly contribute to the CBM and VBM. (d) The band structure of $MAPbBr_3$ (MAPB) with MA along different directions. A direct to indirect band transition is present when the orientation of MA changes from <111> to <001> direction. (e) The Rashba/Dresselhaus effect in the polar phases. The band splitting is present near the CBM and VBM in the polar phase, while in the nonpolar phase, the Rashba/Dresselhaus effect does not exist. (f) The dynamic Rashba effect in MAPI due to the random rotation of MA. Adapted with permission from Ref. [9,24,28,31].

As aforementioned, the phase transition will cause the reorientation of MA cation and change the hydrogen bond, which is very weak and has little contribution to the valence band maximum (VBM), conduction band minimum (CBM) or even band gap (~ 0.1 eV fluctuation) [23,24] (shown in **Fig. 2c**). However, the influence of the collective behavior of MA dynamics on the band structure cannot be neglected, which will influence the photoelectric performance. Liu et al. designed several MA orientations in a supercell and tracked the band gap of MAPI. Their theoretical results showed that the band gap is tunable, ranging from 1.3 to 1.6 eV [25]. Angelis et al. performed ab initio molecular dynamics simulations and also found a variation of ± 0.1 to 0.2 eV of the electronic properties with the ion dynamics [22], which is consistent with Mladenović's works [26]. Besides the value of bandgap, the orientation of MA can also cause the transition from direct band gap to indirect band gap. Sanvito et al. performed van der Waals-corrected DFT calculations and revealed that the bandgap will become indirect if MA orients along a <011>-like direction, which will cause the dynamic change of the band structure and might be the origin of the slow carrier recombination of MAPI [27]. Later they found a similar direct-indirect transition in MAPbBr$_3$ (MAPB). Their DFT calculations demonstrated that MAPB is a direct band-gap semiconductor when MA is oriented along the <111> direction but turns indirect along the <100> direction, as presented in **Fig. 2d** [28].

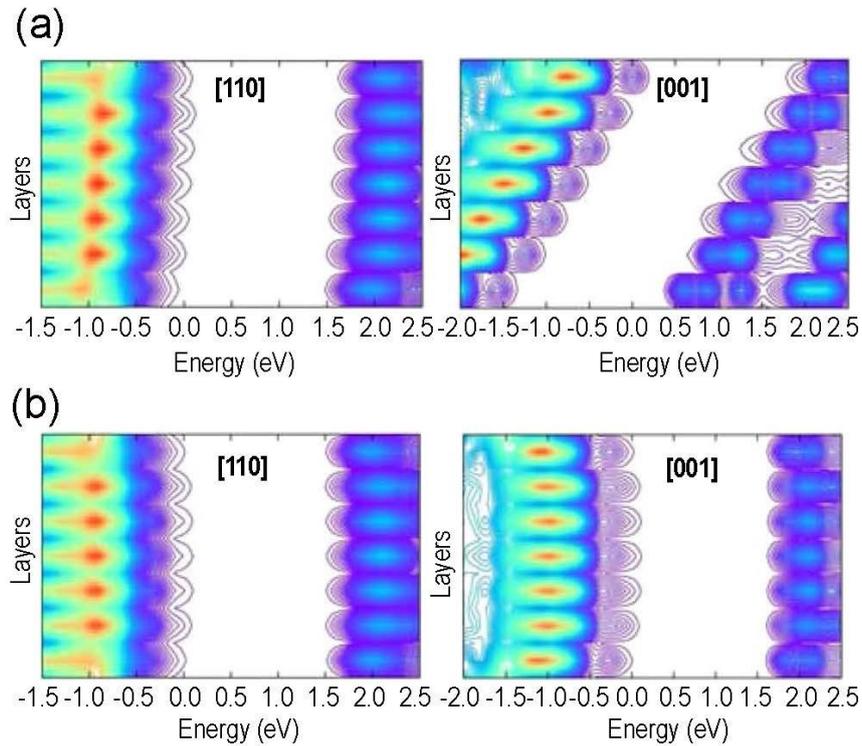

**Figure 3.** The density of states (DOS) of the valence and conduction bands for the surface constructed from (a) polar and (b) nonpolar phases along [110] and [001] directions. Adapted with permission from Ref. [9].

Rashba/Dresselhaus effect is a phenomenon in solid-state physics in which spin-orbit interaction cause energy bands to split, especially present in a crystal system lacking inversion symmetry. The polar OIHP is a typical case to present such effect. In the I4cm polar phase (shown in **Fig. 2a**), the Rashba effect can be detected by ab initio calculations, resulting in splitting of frontier orbitals near fermi level along the M-Γ-Z direction and the creation of indirect band gap as presented in **Fig. 2e** [9,29], while in the I4/mcm nonpolar phase (shown in **Fig. 2b**), the Rashba effect does not exist, as presented in **Fig. 2e** [12]. Fauster et al. used angle-resolved photoelectron spectroscopy and detected the Rashba/Dresselhaus effect in MAPB [30], which is in consistent with theoretical prediction. Furthermore, a 'dynamic Rashba effect' was proposed by Motta et al. due to the rotation of MA or the deformation of the framework when the thermal movement of MA was tracked by van der Waal-corrected ab initio simulations (shown in **Fig. 2f**) [31]. Such effect might lead to the reduced recombination rate due to the spin-forbidden transition [32]. Fauster et al. resonantly excite photocurrents in single-crystalline tetragonal MAPI with circularly polarized light to clarify the existence of such effect. Further studies showed that the energy splitting between the spin-polarized transition and the direct optical transition, as well as the amplitude of the circular photogalvanic effect, increase with temperature [33]. Sum et al. used a broad range of temperature-dependent and time-resolved optical spectroscopies, correlated with density functional theory (DFT) and molecular dynamics (MD) calculations and electrical characterizations and proved the existence of indirect tail states below the direct transition edge in MAPB arising from a dynamical Rashba splitting effect [34]. Recalling the general features of Rashba/Dresselhaus splitting, Kepenekian et al. used symmetry analysis and DFT calculations and discussed the possiblity of designing spintronic devices. They found even in the centrosymmetric system, Rashba effect can also be present under external electric field [35]. The polarization can also impact the electronic properties of the surface structure apart from the bulk. The orientation of MA cations can give rise to strong bending in the valence and conduction bands of polar phases, as exhibited by a gradient in density of states (DOS) along [001] direction in **Fig. 3a**. Such band bending may reduce the carrier recombination and assist charge separation [9]. For nonpolar phase shown in **Fig. 3b**, on the other hand, DOS along both [110] and [001] directions are nearly constant.

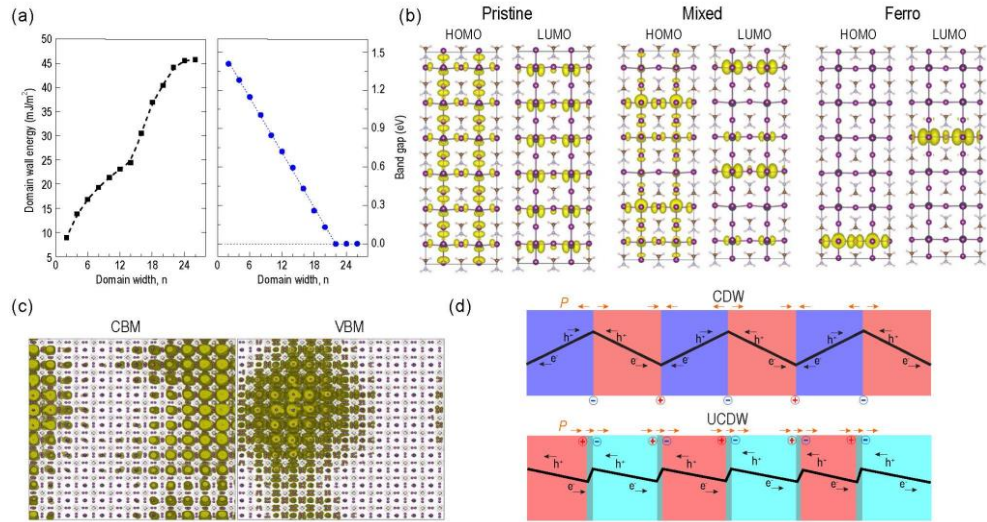

**Figure 4.** Theoretical studies on the properties of the domain wall in MAPI. (a) Domain wall energy and electronic band gap as a function of domain width in MAPI (b) The highest occupied molecular orbital (HOMO) and lowest unoccupied molecular orbital (LUMO) charge densities in pristine MAPI, mixed and ferro systems. (c) The charge density of the CBM and VBM states in the MAPI supercell with MA randomly oriented. Both the CBM and VBM charge densities are strongly localized. (d) The charged (top) and uncharged (bottom) domain walls formed by the orientation of MA cation. Head-to-head and tail-to-tail charge domain walls attract electron and hole, respectively, facilitating the charge separation. Adapted with permission from Refs [36-39].

In the mesoscopic or macroscopic system, the domain wall can be formed in OIHP and display different electronic properties compared with the bulk. Bellaiche et al. studied the formation and bandgap vs the domain width. As shown in **Fig. 4a**, they reported that the domain is stable with rather low formation energy and increasing the domain width can decrease the electronic bandgap from $\approx 1.4$ eV to 0 [36]. The MA orientation can tune the charge aggregation near CBM and VBM [37] (shown in **Fig. 4c**), which might act as the 'ferroelectric highway' and profit the carrier separation [1] The polarization in ferroelectric domains can suppress the nonradiative electron-hole recombination based on the time-domain ab initio study, as shown in **Fig. 4b** [38]. Here pristine system is pure I4cm polar phase with the aligned C-N bonds, mixed system refers to a mixture of aligned and anti-aligned C-N bond pairs, presenting nonpolar characteristics, while ferro-system contains two domains with opposite C-N polar bonds. Charge separation is clearly observed in mixed and ferro-systems with opposite polar axis, beneficial for recombination suppression. Furthermore, when the domain wall is charged, the band gap can be reduced by 20-40%, and there is a strong potential step that facilitates electron-hole separation, as shown in **Fig. 4d**, providing segregated channels for photoexcited charge carriers [39] that is desirable for high conversion efficiency [1] Summarizing all these theoretical studies, there is a general agreement that polarization may be beneficial for photovoltaic conversion.

## EXPERIMENTAL EVIDENCES

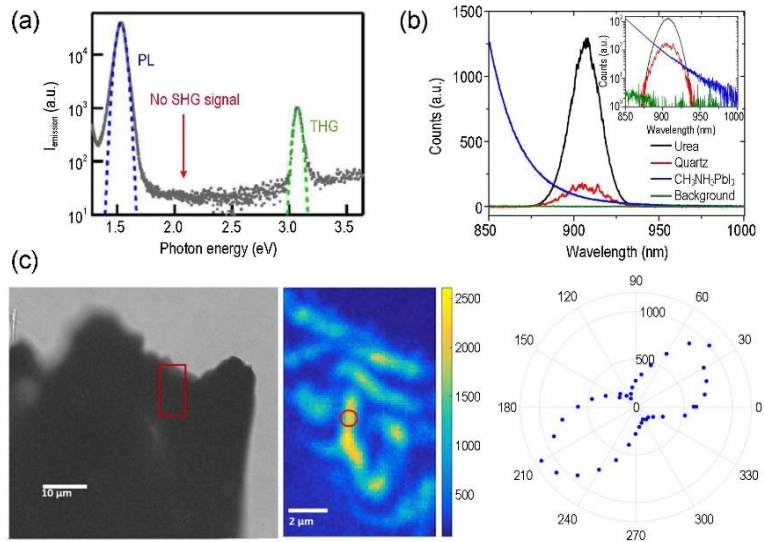

**Figure 5.** Second harmonic generation (SHG) response. (a) Emission spectra of MAPI under excitation at 1.03 eV. (b) Spectra of second harmonic generated at 900 nm, with incident 1800 nm laser pulse measured on Urea (in black) at 1.75 mW incident power, quartz (in red) at 14.9 mW, and MAPI (in blue) at 14.9 mW after subtracting the detector background, which is shown in green. (c) A polar plot of the SHG signal (right) from the marked point in the middle SHG mapping, the area of which is approximately marked by the red box in the bright-field transmission image (left) of a crystal fragment. Adapted with permission from Refs [62,67,76].

Given uncertainty associated with two possible tetragonal lattices for MAPI as shown in **Fig. 1**, it is quite natural to carry out detailed structure analysis via X-ray and neutron diffraction techniques [40-45]. Nevertheless, the subtle structural difference is rather difficult to resolve, and the data can be fit by either polar [20,38,46-58] or nonpolar [8,19,59-68] space groups. Attempts have also been made by transmission electron microscopy (TEM) [69-71], though the materials are prone to degradation [72], and it is virtually impossible to get an atomically resolved image with one exception, wherein HRTEM image acquired from more stable $MAPbBr_3$ (MAPB) shows polar domains [73]. As a result, much effort has been devoted to functional probing, as the properties of polar and nonpolar groups are drastically different. Macroscopic ferroelectric, pyroelectric, and dielectric measurements have also been carried out [48,50,59,60,64,66,74-76], though leakage current and ionic migration often make the data interpretation ambiguous. While polar structure, with the breaking of inversion symmetry, is expected to be active in optical second harmonic generation (SHG) [12], the experimental data are inconclusive as well due to the strong background from other nonlinearities [62,77]. Absence of macroscopic SHG was first reported by Yamada et al. as shown in **Fig. 5a** [67,77], they did not observe any SHG signal under excitation at 1.03 eV (1204 nm) after applying a poling electric field around 1kV/cm to the sample, while third harmonic generation and PL signals are clear due to two-photon absorption. To exclude the possibility that the second-harmonic generated at wavelengths shorter than 800 nm would be strongly absorbed by MAPI in view of its small band gap, Govinda et al. adopted 1800 nm to perform SHG experiments, but the absence of a SHG response at 900 nm is still evident as shown in **Fig. 5b** [62]. It remains possible that ferroelectric domain size is below laser wavelength. Indeed, spatially resolved SHG mapping in **Fig. 5c** provided strong evidence on polar domains in MAPI [76], and local polarity can be averaged out at macroscopic scale, which highlights the importance of spatially resolved local probing.Piezoresponse force microscopy (PFM) is a powerful tool to probe the local electromechanical coupling at the nanoscale [78,79], and it has been widely applied to study OIHPs. Not surprisingly, PFM data reported largely fall into two categories as well, supporting either polar [46,47,49-54,56-58] or nonpolar [8,61,63,65,68,80] structure. In fact, even with quite similar experimental observations, for example characteristic lamellar domain patterns reported by different groups [53,81], the interpretations can be completely opposite. This is because electromechanical responses as probed by PFM can arise from complex microscopic mechanisms [82], especially ionic activities, making PFM data analysis for OIHPs nontrivial. This is best illustrated by the recent debates in Nature Materials [2,3] on the chemical nature of ferroelastic domains in MAPI reported by Liu et al. [68], and there is no agreement on whether it is ferroelectric or not. The latest publication from Liu et al., however, raised an alternative interpretation, that chemical and strain gradients induce flexoelectric polarization in MAPI [83]. This latest study seems to suggest symmetry breaking in MAPI more aligned with polar structure, though its microscopic origin is different.

**Table 2.** Literature survey on the polarity of OIHPs with various compositions.

| Composition | Non-polar | Polar |
| --- | --- | --- |
| $MAPbCl_3$ | Refs [66,86] | Ref. [39] |
| $MAPbBr_3$ | Refs [19,66,87-89] | Refs [39,73,90] |
| $MAPbI_3$ | Refs [5,12,19,42,44,45,59,60,62,68,91] | Refs [6,41,47-58,76,83,92] |
| $FAPbBr_3$ | | Refs [93,94] |
| $FAPbI_3$ | Ref. [95] | |
| $CsPbCl_3$ | Refs [96-98] | |
| $CsPbBr_3$ | Ref. [99] | Refs [100,101] |
| $CsFAMAPbI_xBr_{3-x}$ | Ref. [63] | |

In 2018, we reported an in-depth PFM study [4] on single crystalline MAPI [84], with the goal to resolve the microscopic mechanisms of piezoresponse probed. We acquired the most compelling domain patterns as shown in **Fig. 6a**, and established distinct mechanisms underlying the piezoresponse in adjacent domains, as presented in **Fig. 6b**, suggesting the coexistence of alternating polar and nonpolar structures. In particular, polar domains exhibit predominantly first harmonic linear response that arises from piezoelectricity, while nonpolar domains possesses predominantly second harmonic quadratic response that arises from ionic motion induced electrochemical dipoles [85]. This interpretation is supported by the drastically different thermal variation of piezoresponses in polar and nonpolar domains, one increasing with temperature, the other decreasing with temperature, as shown in **Fig. 6c**, and they converge above cubic-tetragonal transition temperature. When the temperature is reduced, original domain configuration is recovered as seen in **Fig. 6d**, demonstrating strong memory effect. In our view, this set of data unambiguously establish alternating polar and nonpolar domains in our crystal, and this observation can reconcile all the inconsistent experimental data and theoretical analysis reported in literature. Other PFM studies, on the other hand, rarely examine the linear versus quadratic piezoresponses, and thus it is difficult to identify the dominant mechanisms critical for the differentiation. Theoretical calculation suggested that the energetic difference between polar and nonpolar lattice is tiny, less than 100 meV [9], and thus depending on compositions, processing, and environments, the balance can be easily tipped, making both structures possible in experiments. As summarized in **Table 2**, OIHPs with various compositions are reported to be either polar or non-polar [5,6,12,19,39,41,42,44,45,47-60,62-68,73,76,83,86-101]. In addition, the processing methods may also affect the polarity of MAPI. A comparison of the representative preparation methods for MAPI reported to be polar [52-54,57] and non-polar [60,68,102] presents that treating MAPI with dimethylformamide (DMF) vapor on hotplate after general synthesis procedure and inclusion of methylammonium chloride (MACl) or $PbCl_2$, in which chlorine was shown to be beneficial for obtaining MAPI with large grains [103], together with methylammonium iodide (MAI) during synthesis might be favorable for forming polar MAPI. In a sense, we all are both wrong and right, that OIHPs can be both polar and nonpolar.

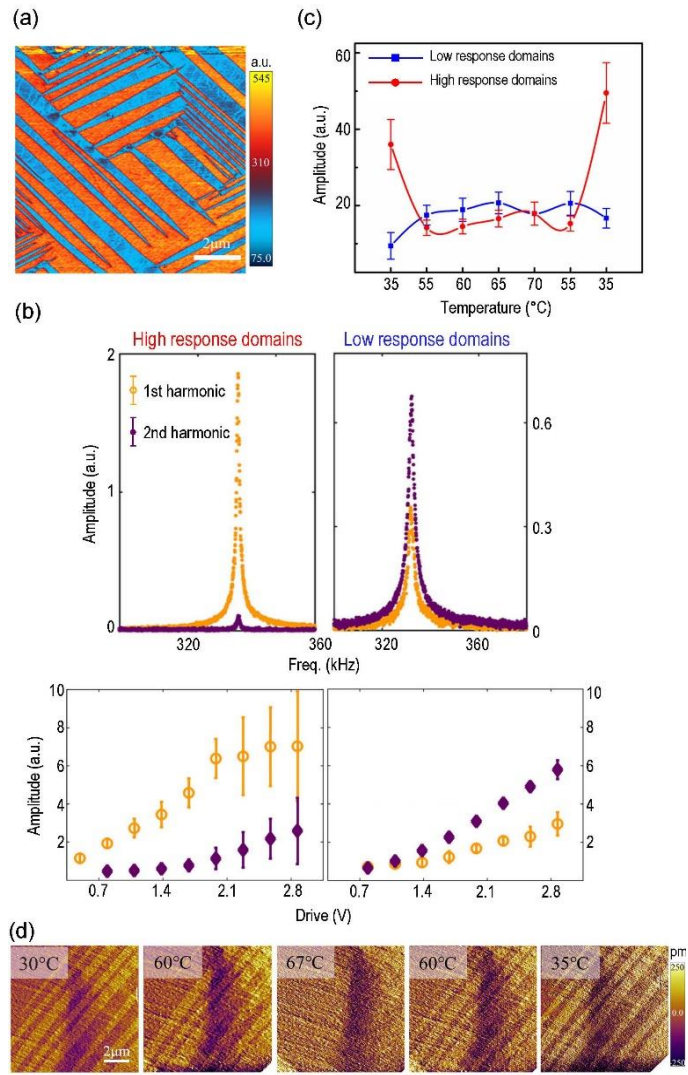

**Figure 6.** Alternating polar and nonpolar domains in MAPI. (a) Ferroic domain patterns of a MAPI crystal revealed by vertical PFM. (b) The first row: point-wise tuning of piezoresponse versus frequency showing a point in high-response polar domain has dominant first harmonic response and negligible second harmonic one, while a point in low-response nonpolar domain has higher second harmonic response; the second row: comparison of first and second harmonic responses versus alternating current voltages averaged over a number of points in high- and low-response domains. (c) Piezoresponses averaged in high-response polar and low-response nonpolar domains showing opposite trend with respect to temperature, yet convergence beyond phase transition. (d) AFM topography mappings under a sequence of temperature across phase transition showing appearance and reemergence of ferroic domains. Adapted with permission from Ref. [4]

If MAPI is polar, can its polarity be switched? Macroscopically this is difficult to resolve, since the data is often smeared by leakage current, ionic migration as well as spatial averaging. Nevertheless, a number of recent reports indicates that electric field can indeed manipulate the domain structures [47,54,104], pointing toward ferroelectric nature of the domain. The unambiguous switching of MAPI domains, however, requires further studies. We also note that ferroelectricity has also been reported in MAPB [90], $CsPbBr_3$ [101], and mixed perovskites [93,105,106].

**PHOTOVOLTAIC IMPLICATIONS**

It is also important to examine the implications of OIHPs' polarity, or lack of it, to the photovoltaic conversion, otherwise the problem remains largely academic. We have indeed observed photo-induced domain switching in MAPI via PFM [58], and similar phenomenon has been observed under photo-excited scanning tunneling microscopy (STM) [107]. The light-domain interactions have been studied by Liu et al. [91,108], and poling has been shown to shift diode characteristic of MAPI [54]. Furthermore, piezoelectric modulations of photocurrent have also been observed [109,110]. All these studies suggest that polar structure may influence photovoltaic conversion process, and to the very least, band bending induced by spontaneous polarization can either promote or hinder carrier transport, depending on its direction. Our study in 2018 indeed revealed that polar domain possesses smaller photocurrent compared to nonpolar one [4] as shown in **Fig. 7a**, and upon heating and cooling across phase transition, memory effect in photocurrent analogue to ferroic domains are also observed, as shown in **Fig. 7b**, confirming modulation of photocurrent by domains.

Nevertheless, there remains a big controversy about the effect of polarization on photovoltaic hysteresis. Unfavorable hysteresis is usually observed in the current-voltage (*I-V*) curve at various scanning rates or directions [111], casting doubts on the validity of the performance of solar cells and making it hard to compare stability data among them. Despite booming research and significant progress on the efficiency of perovskite solar cells, fundamental understanding of frequently observed hysteresis is still inadequate.

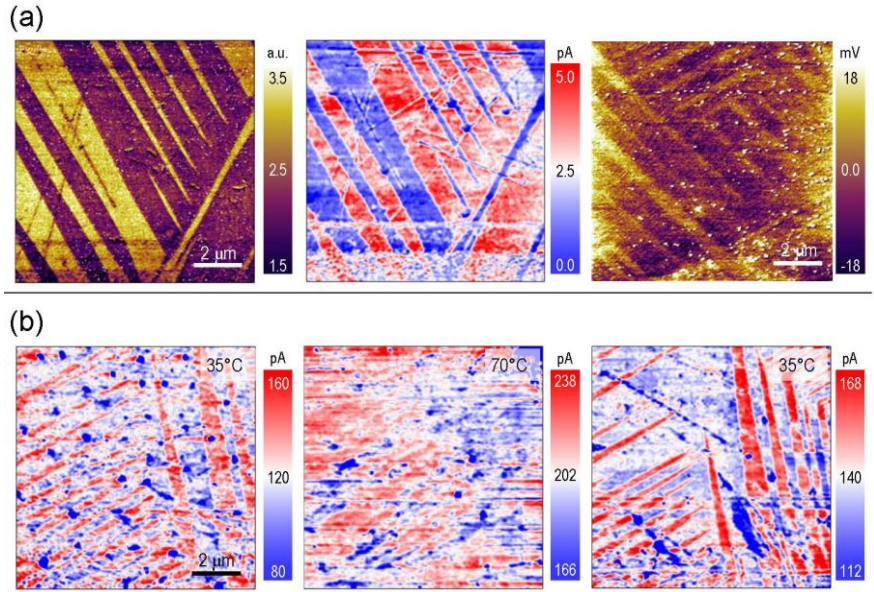

**Figure 7.** Regulation of photocurrent by polar domains. (a) Correlation between PFM (left) and photocurrent (middle) of the same area with reduced photocurrent in polar domains; and surface potential distribution under light follows ferroic domain pattern in **Fig. 6a** with negatively shifted potential in polar domains. (b) Photocurrent distribution in a separate domain pattern at different temperatures across phase transition, showing the disappearing domain pattern at 70 °C upon heating and its reemergence at 35°C after cooling. Adapted with permission from Ref. [4].

Among various interpretations of hysteresis, ferroelectricity was suggested as one plausible origin at the very beginning. For example, Yu et al. attributed the hysteresis to the ferroelectric effect and built a ferroelectric diode model to explain the dependence of hysteresis on the scan range as well as the velocity [112]. They pointed it out that special attention should be paid on the optimization of power conversion efficiency. Recently, Meng et al. investigated the correlations between the interfacial ferroelectricity and the hysteresis of specific heterojunctions by simulations [113]. They found that ferroelectricity is suppressed at the $FAPbI_3/TiO_2$ and $MAPbI_3$/phenyl-C61-butyric-acid-methyl-ester (PCBM) interfaces. The substitution of strong polar MA (dipole moment: 2.29 D) by weak polar FA ions (dipole moment: 0.29 D) and interface passivation could eliminate the interfacial electric field between perovskite and $TiO_2$, leading to consistent interfacial electronic dynamics and the absence of hysteresis [113]. Though it is now generally accepted that ions play a more important role in hysteresis [60,114,115], separation of ionic effect and polar order, however, is not trivial. For example, it has been reported that dipolar Frenkel pair can be induced by ionic migration [116]. Graetzel et al. claimed that hysteresis is due to polarization of halide ion (vacancy) migration in the perovskite layer under the influence of the built-in and applied potential. The mobility of the other possible ionic species ($MA^+$ and $Pb^{2+}$) is much lower and not expected to give any significant contribution in the polarization of devices [114]. We also found that while illumination may enhance polar response in $Cs_{0.05}FA_{0.81}MA_{0.14}PbI_{2.55}Br_{0.45}$ (CsFAMA), only small photovoltaic hysteresis is observed at both nano- and macroscale, demonstrating that even the presence of strong polarization plays negligible role in photovoltaic hysteresis. Based on multi-harmonic measurements, our study support the concept that the primary mechanism responsible for photovoltaic hysteresis is ionic migration instead of polarization for this material [117].

## CONCLUSION AND OUTLOOK

Theoretical calculation is a versatile tool to reveal the interaction of MA with Pb-I framework, study the influence of MA orientation on the optoelectronic properties at nanoscale. Adequate achievements have been reported and some common view has been reached: (*i*) the orientation of MA is determined by the hydrogen bond and usually faces towards low-index direction. (*ii*) the orientation of MA can cause the deformation of the Pb-I framework due to the H···I hydrogen bond, which can break the symmetry of the system and cause the polarization. (*iii*) just tuning the orientation of MA, namely, polarization or not has little influence on the value of band gap, but it can cause the direct-indirect band transition, as well as the Rashba/Dresselhaus effect or even dynamic Rashba/Dresselhaus effect, which may reduce the recombination of carrier, increase the carrier lifetime and enhance the optoelectronic performance.

Experimental observations on the ferroic properties of perovskite solar cells are systematically reviewed, along with its photovoltaic implication: (*i*) subtle difference between polar and nonpolar structure can hardly be resolved by diffraction techniques, TEM, conventional macroscopic measurements, and SHG in a conclusive way due to sample damage or averaging effect; (*ii*) powerful scanning probe can capture spatially resolved functional response from different structures, though caution must be exercised to distinguish complex microscopic mechanisms among polarity, ionic motion, and defect; (*iii*) modulation of photocurrent by polar and nonpolar domains is confirmed, while ions may play a more important role in hysteresis, which is crucial for the performance of solar cells.

At the end of day, we may find that polarization, whatever exact origins, plays just marginal roles in PSCs, but the endeavor often brings in unexpected twists. As shown in **Fig. 8**, giant electrostriction has been reported [116], which was attributed to defect dipoles of Frenkel pairs induced by ionic migration. Here, it seems impossible to distinctly separate ionic migration, defect, and polarity, all of which will be reflected in the experimental observations.

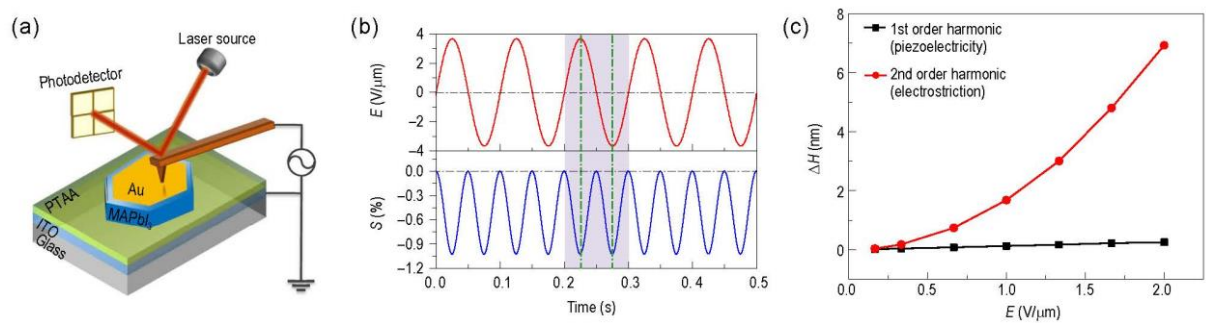

**Figure. 8** Electrostrictive response of MAPI single crystal. (a) Schematic illustration of AFM measurement of strain induced by electric field. (b) Electrostrictive strain of a 40 µm MAPI single crystal under a.c. bias at 10 Hz. (c) Thickness change measured by Mach-Zehnder interferometer due to first-order piezoelectricity and second-order electrostriction for a 40 µm MAPI single crystal under 100 Hz a.c. with different field amplitudes. Adapted with permission from Ref. [116].

Although the beam-induced damage of the synchrotron diffraction and scanning transmission electron microscopy to perovskite samples is inevitable, with the continuous improvement in characterization as well as materials stability [118-121], it is hopeful that these techniques will ultimately settle the debate by resolving the structure details of OIHPs.

For example, Breternitz et al. recently present crystallographic evidence that the symmetry breaking on MAPI comes from the interaction of polar cation MA with the anion framework via synchrotron diffraction [13]. Besides, tentative efforts have been made to mitigate the damage for acquiring atomically resolved imaging [118], including but not limited to using cryo-condition for higher dose tolerance of sample [122], increasing acceleration voltage for decreasing the radiolysis [123], and taking advantages of facet-dependent electron beam sensitivity [119]. Even though further investigation is required to examine their validity, these methods have provided promising direction for the future characterizations. In addition, macroscopic techniques, such as impedance spectroscopy [124,125], in combination with modeling and simulation, may provide valuable supporting data on the microscopic mechanisms. In this regard, integrating local scanning probe with global macroscopic measurements in-situ will provide invaluable microscopic insight into the macroscopic phenomena, which we are trying to develop, and it is particularly important to examine different and often competing dynamical processes from local relaxation studies [126]. From theoretical perspective, as the energy difference between the polar and nonpolar phases is tiny and might varies by different functionals or methods, the calculations with higher accuracy should be performed. In parallel, study on the polarization should be proceeded not only microscopically but also mesoscopically considering the long-range interaction of the ferroelectric domains. Therefore, ab initio calculations of the larger-scale system are also needed. In order to mimic the real experimental conditions, other factors including temperature, strain, and light should also be taken into account to investigate the dynamics of OIHP. Combined with the machine learning and artificial intelligence algorithm [127], the classical molecular dynamic simulation with accurate potential energy surface also needs to be improved. So are OIHPs polar or nonpolar? That might not be the question, but effort in answering it continues to render us new insights.

## ACKNOWLEDGEMENTS

We acknowledge the support of Center for Computational Science and Engineering at Southern University of Science and Technology.

**FUNDING**

This work was supported by the Key-Area Research and Development Program of Guangdong Province (2018B010109009), the Shenzhen Science and Technology Innovation Committee (JCYJ20180507182257563), the National Natural Science Foundation of China (11772207 and 22003074), the Instrument Developing Project of Chinese Academy of Sciences (ZDKYYQ20180004), the Natural Science Foundation of Guangdong Province (2017A030313342 and 2020A1515110580), the State Key Laboratory of Mechanics and Control of Mechanical Structures (Nanjing University of Aeronautics and astronautics) (MCMS-E-0420G01) and the SIAT Innovation Program for Excellent Young Researchers.

*Conflict of interest statement.* None declared.

**REFERENCES**

1. Frost JM, Butler KT and Brivio F *et al.* Atomistic origins of high-performance in hybrid halide perovskite solar cells. *Nano Lett* 2014; **14**: 2584-90.
2. Liu Y, Collins L and Proksch R *et al.* Reply to: on the ferroelectricity of $CH_3NH_3PbI_3$ perovskites. *Nat Mater* 2019; **18**: 1051-3.
3. Schulz AD, Röhm H and Leonhard T *et al.* On the ferroelectricity of $CH_3NH_3 PbI_3$ perovskites. *Nat Mater* 2019; **18**: 1050.
4. Huang B, Kong G and Esfahani EN *et al.* Ferroic domains regulate photocurrent in single-crystalline $CH_3NH_3PbI_3$ films self-grown on $FTO/TiO_2$ substrate. *npj Quantum Mater* 2018; **3**: 30.
5. Baikie T, Fang Y and Kadro JM *et al.* Synthesis and crystal chemistry of the hybrid perovskite $(CH_3NH_3)PbI_3$ for solid-state sensitised solar cell applications. *J Mater Chem A* 2013; **1**: 5628-41.
6. Stoumpos CC, Malliakas CD and Kanatzidis MG. Semiconducting tin and lead iodide perovskites with organic cations: phase transitions, high mobilities, and near-infrared photoluminescent properties. *Inorg Chem* 2013; **52**: 9019-38.
7. Yin WJ, Yang JH and Kang J *et al.* Halide perovskite materials for solar cells: a theoretical review. *J Mater Chem A* 2015; **3**: 8926-42.
8. Fan Z, Xiao J and Sun K *et al.* Ferroelectricity of $CH_3NH_3PbI_3$ perovskite. *J Phys Chem Lett* 2015; **6**: 1155-61.
9. Quarti C, Mosconi E and De Angelis F. Interplay of orientational order and electronic structure in methylammonium lead iodide: implications for solar cell operation. *Chem Mater* 2014; **26**: 6557-69.
10. Stroppa A, Quarti C and De Angelis F *et al.* Ferroelectric polarization of $CH_3NH_3PbI_3$: a detailed study based on density functional theory and symmetry mode analysis. *J Phys Chem Lett* 2015; **6**: 2223-31.

11. Zheng F, Takenaka H and Wang F *et al*. First-principles calculation of the bulk photovoltaic effect in $CH_3NH_3PbI_3$ and $CH_3NH_3PbI_{3-x}Cl_x$. *J Phys Chem Lett* 2015; **6**: 31-7.
12. Frohna K, Deshpande T and Harter J *et al*. Inversion symmetry and bulk Rashba effect in methylammonium lead iodide perovskite single crystals. *Nat Commun* 2018; **9**: 1829.
13. Breternitz J, Lehmann F and Barnett SA *et al*. Role of the iodide-methylammonium interaction in the ferroelectricity of $CH_3NH_3PbI_3$. *Angew Chem Int Ed* 2020; **59**: 424-8.
14. Brivio F, Walker AB and Walsh A. Structural and electronic properties of hybrid perovskites for high-efficiency thin-film photovoltaics from first-principles. *APL Mater* 2013; **1**: 042111.
15. Bechtel JS, Seshadri R and Van der Ven A. Energy landscape of molecular motion in cubic methylammonium lead iodide from first-principles. *J Phys Chem C* 2016; **120**: 12403-10.
16. Shimamura K, Hakamata T and Shimojo F *et al*. Rotation mechanism of methylammonium molecules in organometal halide perovskite in cubic phase: an ab initio molecular dynamics study. *J Chem Phys* 2016; **145**: 224503.
17. Xu Q, Stroppa A and Lv J *et al*. Impact of organic molecule rotation on the optoelectronic properties of hybrid halide perovskites. *Phys Rev Mater* 2019; **3**: 125401.
18. Svane KL, Forse AC and Grey CP *et al*. How strong is the hydrogen bond in hybrid perovskites? *J Phys Chem Lett* 2017; **8**: 6154-9.
19. Govinda S, Kore BP and Bokdam M *et al*. Behavior of methylammonium dipoles in $MAPbX_3$ (X = Br and I). *J Phys Chem Lett* 2017; **8**: 4113-21.
20. Li Y, Behtash M and Wong J *et al*. Enhancing ferroelectric dipole ordering in organic-inorganic hybrid perovskite $CH_3NH_3PbI_3$: strain and doping engineering. *J Phys Chem C* 2018; **122**: 177-84.
21. Mattoni A, Filippetti A and Saba M *et al*. Methylammonium rotational dynamics in lead halide perovskite by classical molecular dynamics: the role of temperature. *J Phys Chem C* 2015; **119**: 17421-8.
22. Mosconi E, Quarti C and Ivanovska T *et al*. Structural and electronic properties of organo-halide lead perovskites: a combined IR-spectroscopy and ab initio molecular dynamics investigation. *Phys Chem Chem Phys* 2014; **16**: 16137-44.
23. Carignano MA, Kachmar A and Hutter Jr. Thermal effects on $CH_3NH_3PbI_3$ perovskite from ab initio molecular dynamics simulations. *J Phys Chem C* 2015; **119**: 8991-7.
24. Long R, Fang W and Prezhdo OV. Moderate humidity delays electron-hole recombination in hybrid organic-inorganic perovskites: time-domain ab initio simulations rationalize experiments. *J Phys Chem Lett* 2016; **7**: 3215-22.
25. Geng W, Hu Q and Tong CJ *et al*. The Influence of dipole moments induced by organic molecules and domain structures on the properties of $CH_3NH_3PbI_3$ perovskite. *Adv Theory Simul* 2019; **2**: 1900041.
26. Mladenović M and Vukmirović N. Effects of thermal disorder on the electronic structure of halide perovskites: insights from MD simulations. *Phys Chem Chem Phys* 2018; **20**: 25693-700.
27. Motta C, El-Mellouhi F and Kais S *et al*. Revealing the role of organic cations in hybrid halide perovskite $CH_3NH_3PbI_3$. *Nat Commun* 2015; **6**: 7026.
28. Motta C, El-Mellouhi F and Sanvito S. Exploring the cation dynamics in lead-bromide hybrid perovskites. *Phys Rev B* 2016; **93**: 235412.
29. Leppert L, Reyes-Lillo SE and Neaton JB. Electric field-and strain-induced Rashba effect in hybrid halide perovskites. *J Phys Chem Lett* 2016; **7**: 3683-9.

30. Niesner D, Wilhelm M and Levchuk I *et al.* Giant Rashba splitting in $CH_3NH_3PbBr_3$ organic-inorganic perovskite. *Phys Rev Lett* 2016; **117**: 126401.

31. Etienne T, Mosconi E and De Angelis F. Dynamical origin of the Rashba effect in organohalide lead perovskites: a key to suppressed carrier recombination in perovskite solar cells? *J Phys Chem Lett* 2016; **7**: 1638-45.

32. Zheng F, Tan LZ and Liu S *et al.* Rashba spin-orbit coupling enhanced carrier lifetime in $CH_3NH_3PbI_3$. *Nano Lett* 2015; **15**: 7794-800.

33. Niesner D, Hauck M and Shrestha S *et al.* Structural fluctuations cause spin-split states in tetragonal $(CH_3NH_3)PbI_3$ as evidenced by the circular photogalvanic effect. *Proc Natl Acad Sci USA* 2018; **115**: 9509-14.

34. Wu B, Yuan HF and Xu Q *et al.* Indirect tail states formation by thermal-induced polar fluctuations in halide perovskites. *Nat Commun* 2019; **10**: 484.

35. Kepenekian M, Robles R and Katan C *et al.* Rashba and Dresselhaus effects in hybrid organic-inorganic perovskites: from basics to devices. *ACS Nano* 2015; **9**: 11557-67.

36. Chen L, Paillard C and Zhao HJ *et al.* Tailoring properties of hybrid perovskites by domain-width engineering with charged walls. *npj Comput Mater* 2018; **4**: 75.

37. Ma J and Wang L-W. Nanoscale charge localization induced by random orientations of organic molecules in hybrid perovskite $CH_3NH_3PbI_3$. *Nano Lett* 2015; **15**: 248-53.

38. Qiao L, Fang W-H and Long R. Ferroelectric polarization suppresses nonradiative electron-hole recombination in $CH_3NH_3PbI_3$ perovskites: a time-domain ab initio study. *J Phys Chem Lett* 2019; **10**: 7237-44.

39. Liu S, Zheng F and Koocher NZ *et al.* Ferroelectric domain wall induced band gap reduction and charge separation in organometal halide perovskites. *J Phys Chem Lett* 2015; **6**: 693-9.

40. Baikie T, Barrow NS and Fang YA *et al.* A combined single crystal neutron/X-ray diffraction and solid-state nuclear magnetic resonance study of the hybrid perovskites $CH_3NH_3PbX_3$ (X = I, Br and Cl). *J Mater Chem A* 2015; **3**: 9298-307.

41. Dang YY, Liu Y and Sun YX *et al.* Bulk crystal growth of hybrid perovskite material $CH_3NH_3PbI_3$. *Crystengcomm* 2015; **17**: 665-70.

42. Fang HH, Raissa R and Abdu-Aguye M *et al.* Photophysics of organic-inorganic hybrid lead iodide perovskite single crystals. *Adv Funct Mater* 2015; **25**: 2378-85.

43. Sewvandi GA, Kodera K and Ma H *et al.* Antiferroelectric nature of $CH_3NH_3PbI_{3-x}Cl_x$ perovskite and its implication for charge separation in perovskite solar cells. *Sci Rep* 2016; **6**: 30680.

44. Weller MT, Weber OJ and Henry PF *et al.* Complete structure and cation orientation in the perovskite photovoltaic methylammonium lead iodide between 100 and 352 K. *Chem Commun* 2015; **51**: 4180-3.

45. Whitfield PS, Herron N and Guise WE *et al.* Structures, phase transitions and tricritical behavior of the hybrid perovskite methyl ammonium lead iodide. *Sci Rep* 2016; **6**: 35685.

46. Chen B, Shi J and Zheng X *et al.* Ferroelectric solar cells based on inorganic-organic hybrid perovskites. *J Mater Chem A* 2015; **3**: 7699-705.

47. Garten LM, Moore DT and Nanayakkara SU *et al.* The existence and impact of persistent ferroelectric domains in $MAPbI_3$. *Sci Adv* 2019; **5**: eaas9311.

48. Guo H, Liu P and Zheng S *et al.* Re-entrant relaxor ferroelectricity of methylammonium lead iodide. *Curr Appl Phys* 2016; **16**: 1603-6.

49. Kim H-S, Kim SK and Kim BJ *et al.* Ferroelectric polarization in $CH_3NH_3PbI_3$ perovskite. *J Phys Chem Lett* 2015; **6**: 1729-35.
50. Kim Y-J, Dang T-V and Choi H-J *et al.* Piezoelectric properties of $CH_3NH_3PbI_3$ perovskite thin films and their applications in piezoelectric generators. *J Mater Chem A* 2016; **4**: 756-63.
51. Kutes Y, Ye L and Zhou Y *et al.* Direct observation of ferroelectric domains in solution-processed $CH_3NH_3PbI_3$ perovskite thin films. *J Phys Chem Lett* 2014; **5**: 3335-9.
52. Leonhard T, Schulz AD and Röhm H *et al.* Probing the microstructure of methylammonium lead iodide perovskite solar cells. *Energy Technol* 2019; **7**: 1800989.
53. Röhm H, Leonhard T and Hoffmann M *et al.* Ferroelectric domains in methylammonium lead iodide perovskite thin-films. *Energy Environ Sci* 2017; **10**: 950-5.
54. Röhm H, Leonhard T and Hoffmann MJ *et al.* Ferroelectric poling of methylammonium lead iodide thin films. *Adv Funct Mater* 2019; **30**: 1908657.
55. Rohm H, Leonhard T and Schulz AD *et al.* Ferroelectric properties of perovskite thin films and their implications for solar energy conversion. *Adv Mater* 2019; **31**: 1806661.
56. Seol D, Jeong A and Han MH *et al.* Origin of hysteresis in $CH_3NH_3PbI_3$ perovskite thin films. *Adv Funct Mater* 2017; **27**: 1701924.
57. Vorpahl SM, Giridharagopal R and Eperon GE *et al.* Orientation of ferroelectric domains and disappearance upon heating methylammonium lead triiodide perovskite from tetragonal to cubic phase. *ACS Appl Energy Mater* 2018; **1**: 1534-9.
58. Wang P, Zhao J and Wei L *et al.* Photo-induced ferroelectric switching in perovskite $CH_3NH_3PbI_3$ films. *Nanoscale* 2017; **9**: 3806-17.
59. Anusca I, Balčiūnas S and Gemeiner P *et al.* Dielectric response: answer to many questions in the methylammonium lead halide solar cell absorbers. *Adv Energy Mater* 2017; **7**: 1700600.
60. Beilsten-Edmands J, Eperon G and Johnson R *et al.* Non-ferroelectric nature of the conductance hysteresis in $CH_3NH_3PbI_3$ perovskite-based photovoltaic devices. *Appl Phys Lett* 2015; **106**: 173502.
61. Coll M, Gomez As and Mas-Marza E *et al.* Polarization switching and light-enhanced piezoelectricity in lead halide perovskites. *J Phys Chem Lett* 2015; **6**: 1408-13.
62. G S, Mahale P and Kore BP *et al.* Is $CH_3NH_3PbI_3$ polar? *J Phys Chem Lett* 2016; **7**: 2412-9.
63. Gómez A, Wang Q and Goñi AR *et al.* Ferroelectricity-free lead halide perovskites. *Energy Environ Sci* 2019; **12**: 2537-47.
64. Hoque MNF, Yang M and Li Z *et al.* Polarization and dielectric study of methylammonium lead iodide thin film to reveal its nonferroelectric nature under solar cell operating conditions. *ACS Energy Lett* 2016; **1**: 142-9.
65. Liu Y, Collins L and Belianinov A *et al.* Dynamic behavior of $CH_3NH_3PbI_3$ perovskite twin domains. *Appl Phys Lett* 2018; **113**: 072102.
66. Šimėnas M, Balčiūnas S and Mączka M *et al.* Exploring the antipolar nature of methylammonium lead halides: a Monte Carlo and pyrocurrent study. *J Phys Chem Lett* 2017; **8**: 4906-11.
67. Yamada Y, Yamada T and Phuong LQ *et al.* Dynamic optical properties of $CH_3NH_3PbI_3$ single crystals as revealed by one-and two-photon excited photoluminescence measurements. *J Am Chem Soc* 2015; **137**: 10456-9.
68. Liu Y, Collins L and Proksch R *et al.* Chemical nature of ferroelastic twin domains in $CH_3NH_3PbI_3$ perovskite. *Nat Mater* 2018; **17**: 1013-9.

69. Chen S, Zhang X and Zhao J *et al.* Atomic scale insights into structure instability and decomposition pathway of methylammonium lead iodide perovskite. *Nat Commun* 2018; **9**: 4807.

70. Rothmann MU, Li W and Zhu Y *et al.* Direct observation of intrinsic twin domains in tetragonal $CH_3NH_3PbI_3$. *Nat Commun* 2017; **8**: 14547.

71. Rothmann MU, Li W and Zhu Y *et al.* Structural and chemical changes to $CH_3NH_3PbI_3$ induced by electron and gallium ion beams. *Adv Mater* 2018; **30**: 1800629.

72. Saliba M, Stolterfoht M and Wolff CM *et al.* Measuring aging stability of perovskite solar cells. *Joule* 2018; **2**: 1019-24.

73. Zhang D, Zhu Y and Liu L *et al.* Atomic-resolution transmission electron microscopy of electron beam-sensitive crystalline materials. *Science* 2018; **359**: 675-9.

74. Birkhold ST, Hu H and Höger PT *et al.* Mechanism and impact of cation polarization in methylammonium lead iodide. *J Phys Chem C* 2018; **122**: 12140-7.

75. Cordero F, Craciun F and Trequattrini F *et al.* Competition between polar and antiferrodistortive modes and correlated dynamics of the methylammonium molecules in $MAPbI_3$ from anelastic and dielectric measurements. *J Phys Chem Lett* 2018; **9**: 4401-6.

76. Rakita Y, Bar-Elli O and Meirzadeh E *et al.* Tetragonal $CH_3NH_3PbI_3$ is ferroelectric. *Proc Natl Acad Sci USA* 2017; **114**: E5504-12.

77. Govinda S, Kore BP and Mahale P *et al.* Can SHG measurements determine the polarity of hybrid lead halide perovskites? *ACS Energy Lett* 2018; **3**: 1887-91.

78. Bonnell DA, Kalinin SV and Kholkin A *et al.* Piezoresponse force microscopy: a window into electromechanical behavior at the nanoscale. *MRS Bull* 2009; **34**: 648-57.

79. Li J, Li J-F and Yu Q *et al.* Strain-based scanning probe microscopies for functional materials, biological structures, and electrochemical systems. *J Materiomics* 2015; **1**: 3-21.

80. Hermes IM, Bretschneider SA and Bergmann VW *et al.* Ferroelastic fingerprints in methylammonium lead iodide perovskite. *J Phys Chem C* 2016; **120**: 5724-31.

81. Strelcov E, Dong Q and Li T *et al.* $CH_3NH_3PbI_3$ perovskites: ferroelasticity revealed. *Sci Adv* 2017; **3**: e1602165.

82. Chen QN, Ou Y and Ma F *et al.* Mechanisms of electromechanical coupling in strain based scanning probe microscopy. *Appl Phys Lett* 2014; **104**: 242907.

83. Liu Y, Ievlev AV and Collins L *et al.* Strain-chemical gradient and polarization in metal halide perovskites. *Adv Electron Mater* 2020; **6**: 1901235.

84. Zhao J, Kong G and Chen S *et al.* Single crystalline $CH_3NH_3PbI_3$ self-grown on $FTO/TiO_2$ substrate for high efficiency perovskite solar cells. *Sci Bull* 2017; **62**: 1173-6.

85. Yu J, Esfahani EN and Zhu Q *et al.* Quadratic electromechanical strain in silicon investigated by scanning probe microscopy. *J Appl Phys* 2018; **123**: 155104.

86. Bari M, Bokov AA and Ye ZG. Ferroelasticity, domain structures and phase symmetries in organic-inorganic hybrid perovskite methylammonium lead chloride. *J Mater Chem C* 2020; **8**: 9625-31.

87. Gonzalez-Carrero S, Frances-Soriano L and Gonzalez-Bejar M *et al.* The luminescence of $CH_3NH_3PbBr_3$ perovskite nanoparticles crests the summit and their photostability under wet conditions is enhanced. *Small* 2016; **12**: 5245-50.

88. Mashiyama H, Kawamura Y and Kubota Y. The anti-polar structure of $CH_3NH_3PbBr_3$. *J Korean Phys Soc* 2007; **51**: 850-3.

89. Bari M, Bokov AA and Ye Z-G. Ferroelastic domains and phase transitions in organic–inorganic hybrid perovskite $CH_3NH_3PbBr_3$. *J Mater Chem C* 2021; **9**: 3096-107.

90. Gao Z-R, Sun X-F and Wu Y-Y *et al.* Ferroelectricity of the orthorhombic and tetragonal $MAPbBr_3$ single crystal. *J Phys Chem Lett* 2019; **10**: 2522-7.

91. Liu Y, Ievlev AV and Collins L *et al.* Light-ferroic interaction in hybrid organic-inorganic perovskites. *Adv Opt Mater* 2019; **7**: 1901451.

92. Frost JM, Butler KT and Walsh A. Molecular ferroelectric contributions to anomalous hysteresis in hybrid perovskite solar cells. *APL Mater* 2014; **2**: 081506.

93. Ahmadi M, Collins L and Puretzky A *et al.* Exploring anomalous polarization dynamics in organometallic halide perovskites. *Adv Mater* 2018; **30**: 1705298.

94. Ding R, Liu H and Zhang XL *et al.* Flexible piezoelectric nanocomposite generators based on formamidinium lead halide perovskite nanoparticles. *Adv Funct Mater* 2016; **26**: 7708-16.

95. Taylor VCA, Tiwari D and Duchi M *et al.* Investigating the role of the organic cation in formamidinium lead iodide perovskite using ultrafast spectroscopy. *J Phys Chem Lett* 2018; **9**: 895-901.

96. Voloshinovskii A, Myagkota S and Levitskii R. Luminescence of ferroelastic $CsPbCl_3$ nanocrystals. *Ferroelectrics* 2005; **317**: 311-5.

97. Lim AR and Jeong S-Y. Ferroelastic phase transition and twin structure by $^{133}Cs$ NMR in a $CsPbCl_3$ single crystal. *Physica B Condens Matter* 2001; **304**: 79-85.

98. Hirotsu S. Experimental studies of structural phase transitions in $CsPbCl_3$. *J Phys Soc Japan* 1971; **31**: 552-60.

99. Marçal LA, Oksenberg E and Dzhigaev D *et al. In situ* imaging of ferroelastic domain dynamics in $CsPbBr_3$ perovskite nanowires by nanofocused scanning X-ray diffraction. *ACS Nano* 2020; **14**: 15973-82.

100. Zhao Y-Q, Ma Q-R and Liu B *et al.* Pressure-induced strong ferroelectric polarization in tetra-phase perovskite $CsPbBr_3$. *Phys Chem Chem Phys* 2018; **20**: 14718-24.

101. Li X, Chen S and Liu P-F *et al.* Evidence for ferroelectricity of all-inorganic perovskite $CsPbBr_3$ quantum dots. *J Am Chem Soc* 2020; **142**: 3316-20.

102. Chen Q, Zhou HP and Hong ZR *et al.* Planar heterojunction perovskite solar cells via vapor-assisted solution process. *J Am Chem Soc* 2014; **136**: 622-5.

103. Luo SQ and Daoud WA. Crystal structure formation of $CH_3NH_3PbI_{3-x}Cl_x$ perovskite. *Materials* 2016; **9**: 123.

104. Kim D, Yun JS and Sharma P *et al.* Light-and bias-induced structural variations in metal halide perovskites. *Nat Commun* 2019; **10**: 444.

105. Xiao J, Chang J and Li B *et al.* Room temperature ferroelectricity of hybrid organic-inorganic perovskites with mixed iodine and bromine. *J Mater Chem A* 2018; **6**: 9665-76.

106. Zhang Q, Solanki A and Parida K *et al.* Tunable ferroelectricity in Ruddlesden-Popper halide perovskites. *ACS Appl Mater Interfaces* 2019; **11**: 13523-32.

107. Hsu H-C, Huang B-C and Chin S-C *et al.* Photodriven dipole reordering: key to carrier separation in metalorganic halide perovskites. *ACS Nano* 2019; **13**: 4402-9.

108. Liu Y, Li M and Wang M *et al.* Twin domains modulate light-matter interactions in metal halide perovskites. *APL Mater* 2020; **8**: 011106.

109. Khorramshahi F, Woughter AG and Ram MK *et al.* Apparent piezo-Energy & Environmental Sciencephotocurrent modulation in methylammonium lead iodide perovskite photodetectors. *Adv Electron Mater* 2019; **5**: 1900518.

110. Lai Q, Zhu L and Pang Y *et al.* Piezo-phototronic effect enhanced photodetector based on $CH_3NH_3PbI_3$ single crystals. *ACS Nano* 2018; **12**: 10501-8.

111. Ravishankar S, Gharibzadeh S and Roldan-Carmona C *et al.* Influence of charge transport layers on open-circuit voltage and hysteresis in perovskite solar cells. *Joule* 2018; **2**: 788-98.

112. Wei J, Zhao Y and Li H *et al.* Hysteresis analysis based on the ferroelectric effect in hybrid perovskite solar cells. *J Phys Chem Lett* 2014; **5**: 3937-45.

113. Ma W, Zhang X and Xu Z *et al.* Reducing anomalous hysteresis in perovskite solar cells by suppressing the interfacial ferroelectric order. *ACS Appl Mater Interfaces* 2020; **12**: 12275-84.

114. Meloni S, Moehl T and Tress W *et al.* Ionic polarization-induced current-voltage hysteresis in $CH_3NH_3PbX_3$ perovskite solar cells. *Nat Commun* 2016; **7**: 10334.

115. Shi JJ, Li YM and Li YS *et al.* From ultrafast to ultraslow: charge-carrier dynamics of perovskite solar cells. *Joule* 2018; **2**: 879-901.

116. Chen B, Li T and Dong Q *et al.* Large electrostrictive response in lead halide perovskites. *Nat Mater* 2018; **17**: 1020-6.

117. Xia G, Huang B and Zhang Y *et al.* Nanoscale insights into photovoltaic hysteresis in triple-cation mixed-halide perovskite: resolving the role of polarization and ionic migration. *Adv Mater* 2019; **31**: 1902870.

118. Chen S and Gao P. Challenges, myths, and opportunities of electron microscopy on halide perovskites. *J Appl Phys* 2020; **128**: 010901.

119. Chen S, Zhang Y and Zhao J *et al.* Transmission electron microscopy of organic-inorganic hybrid perovskites: myths and truths. *Sci Bull* 2020; **65**: 1643-9.

120. Ono LK, Qi YB and Liu SZ. Progress toward stable lead halide perovskite solar cells. *Joule* 2018; **2**: 1961-90.

121. Zhou YY, Sternlicht H and Padture NP. Transmission electron microscopy of halide perovskite materials and devices. *Joule* 2019; **3**: 641-61.

122. Li Y, Zhou W and Li Y *et al.* Unravelling degradation mechanisms and atomic structure of organic-inorganic halide perovskites by cryo-EM. *Joule* 2019; **3**: 2854-66.

123. Dang Z, Shamsi J and Palazon F *et al. In situ* transmission electron microscopy study of electron beam-induced transformations in colloidal cesium lead halide perovskite nanocrystals. *ACS Nano* 2017; **11**: 2124-32.

124. Jeong S-H, Park J and Han T-H *et al.* Characterizing the efficiency of perovskite solar cells and light-emitting diodes. *Joule* 2020; **4**: 1206-35.

125. Pitarch-Tena D, Ngo TT and Vallés-Pelarda M *et al.* Impedance spectroscopy measurements in perovskite solar cells: device stability and noise reduction. *ACS Energy Lett* 2018; **3**: 1044-8.

126. Yu JX, Huang BY and Li AL *et al.* Resolving local dynamics of dual ions at the nanoscale in electrochemically active materials. *Nano Energy* 2019; **66**: 104160.

127. Correa-Baena JP, Hippalgaonkar K and van Duren J *et al.* Accelerating materials development via automation, machine learning, and high-performance computing. *Joule* 2018; **2**: 1410-20.